# A phase-stable dual-comb interferometer

Zaijun Chen[1,2], Ming Yan[1,2], Theodor W. Hänsch[1,2], Nathalie Picqué[1,2,*]
1. Max-Planck-Institut für Quantenoptik, Hans-Kopfermann-Straße 1, 85748 Garching, Germany
2. Ludwig-Maximilians-Universität München, Fakultät für Physik, Schellingstr. 4/III, 80799 München, Germany
* corresponding author: nathalie.picque@mpq.mpg.de

**Abstract**

Improvements to dual-comb interferometers will benefit precision spectroscopy and sensing[1-4], distance metrology[5], tomography[6], telecommunications[7] etc. A specific requirement of such interferometers is to enforce mutual coherence between the two combs[8] over the measurement time. With feed-forward relative stabilization of the carrier-enveloppe offset frequencies, we experimentally realize such mutual coherence over times that exceed 300 seconds, two orders of magnitude longer than state-of-the-art systems. Illustration is given with near-infrared Fourier transform molecular spectroscopy, where two combs of slightly different repetition frequencies replace a scanning two-beam interferometer. Our technique can be implemented with any frequency comb generators including microresonators[9-12] or quantum cascade lasers[13].

The precise control of the phase difference in a two-beam interferometer involving a moveable mirror has been perfected over decades and the mutual coherence between the two arms of the interferometer can be maintained over tens of hours [14] in standard laboratory environments. With a dual-comb system, a type of two-beam interferometer where, in most of the implementations, the phase difference is automatically and periodically scanned by means of two asynchronous trains of pulses, it is still challenging to control the relative timing and phase fluctuations between the two combs. Such systems have a potential for precisions directly set by atomic clocks though. The most powerful approach for establishing mutual coherence between two frequency combs has been to lock each comb to the same pair of cavity-stabilized continuous-wave lasers with Hz-level linewidth. In this way, mutual coherence times of the order of 1 s, determined by the linewidth of the continuous-wave lasers, have been achieved and linear-phase correction enhances the effective averaging times to tens of minutes [4]. Alternatively, schemes correcting the relative fluctuations with analog electronics [3], digital processing [15] or computer algorithms [16] permit measurements, even with free-running lasers. Another current trend is to design systems with built-in passive mutual coherence [2,17]. None of these solutions reaches the overall performance of the cavity-locked systems. With the increasing number of foreseen applications for highly-precise dual-comb systems, for instance to spectroscopic measurements of very weak lines, to Doppler-free broadband spectroscopy [18], to precise measurements of refractive indexes [19] or to distance monitoring [20] between formations of spacecrafts, breaking the barrier of 1 second for the interferometer coherence times is crucial.

We propose and implement a technique of dual-comb interferometry which demonstrates mutual coherence times of 290 seconds, without any indication that a limit is reached. We use feed-forward adjustment of the relative carrier-enveloppe offset frequencies of the two combs with an external actuator which permits very fast response time without locking electronics and may be used on any types of frequency comb generators. Feed-forward control of a laser is a technique known for fast response time and low noise, as successfully demonstrated for the frequency stabilization of continous-wave lasers [21], for the stabilization of a mode-locked laser to a Fabry-Perot cavity [22] or for carrier-envelope offset lock of frequency combs [23]. In our scheme (Fig.1), one frequency comb (master), of repetition frequency $f_{rep}$ and carrier-envelope offset $f_{ceo}$, is stabilized against a radio-frequency clock using the traditional self-referenced scheme, providing high frequency accuracy over long time scales. The second comb generator (slave), of slightly different repetition frequency $f_{rep}+\delta f_{rep}$, with $\delta f_{rep}$ small compared to $f_{rep}$, and of carrier-envelope offset $f_{ceo}+\delta f_{ceo}$, follows the rapidly varying instabilities of the first master comb. Two beat-notes, each between one line of the master comb and one line of the slave comb, serve as indicators of the relative fluctuations between the combs and are maintained at fixed frequency offsets. One beatnote generates the error signal for feeding an acousto-optic frequency shifter at the output of the slave comb: all the spectral lines in the first-order diffracted beam of the slave comb are shifted by the same amount. Relative carrier-envelope offset control is therefore achieved with a bandwidth on the MHz-level. The second beatnote locks the repetition frequency of the slave comb through a slow feedback loop, which translates one of the cavity mirror of the slave comb by means of a piezo-electric transducer. The choice of the actuators and of the way to generate the error signals may be adapted to the type of employed frequency comb generators: for instance, combining an electro-optic phase modulator and an acousto-optic frequency shifter may lead to faster response times.

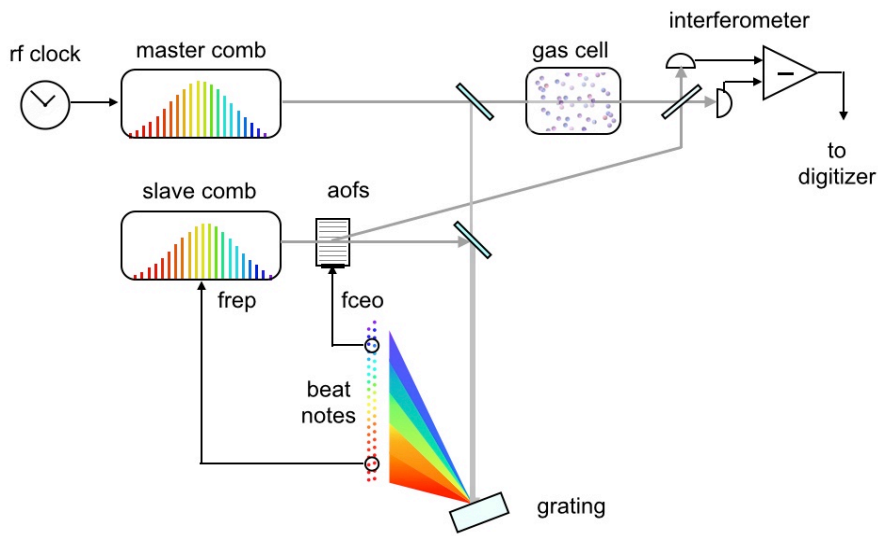

**Figure 1. Feed-forward dual-comb spectroscopy.** A self-referenced master comb provides long term stability and the slave frequency comb follows the drifts of comb 1, providing high-bandwidth mutual coherence. The beam of the master comb interrogates the sample and beats with the beam diffracted in the first-order by the acousto-optic frequency shifter. The optical signal is detected with a balanced differential detector and is digitized.

We illustrate the performance of our interferometer with a set-up dedicated to near-infrared dual-comb spectroscopy (Fig. 1). Two commercial femtosecond erbium-doped amplified fiber lasers emitting around 190 THz are used. Their repetition frequencies are such that $f_{rep}$=100 MHz and $\delta f_{rep}$= 100 Hz. The outputs of the combs are spectrally broadened in highly nonlinear fibers to span the region from 166 to 245 THz. The master comb interrogates a single-pass cell filled with a gas at low pressure, while the first-order diffracted beam of the slave comb serves as local oscillator. The two combs are combined on a beam-mixer. For improved signal-to-noise ratio in a selected spectral region, the optical signal may be spectrally filtered with home-made grating filters of tunable central wavelength and spectral width. The two outputs of the interferometer are detected by a differential detector. The time-domain interference signal is digitized with a data acquisition board.

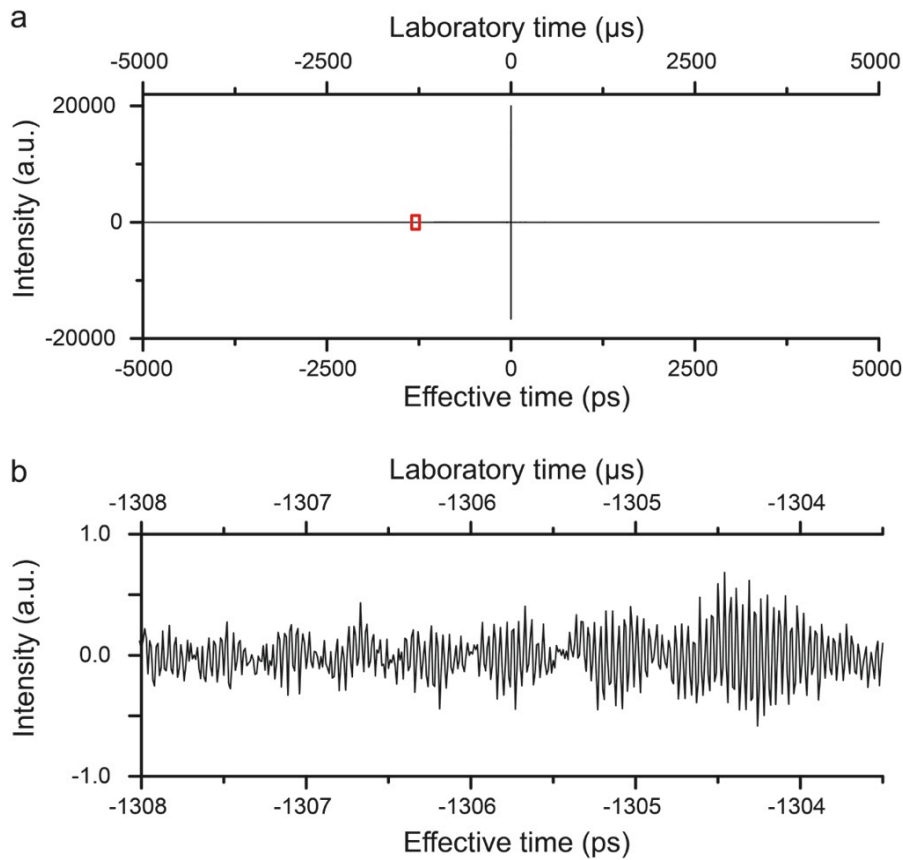

**Figure 2. Time domain interferogram.** In the laboratory time frame, the interferograms repeat with a period of $1/\delta f_{rep}$=10 ms, which corresponds to optical delays of $1/f_{rep}$=10 ns. **a.** 29000 consecutive interferograms are averaged in the time domain over a span of 10 ms, resulting in a total laboratory time of 290 s. **b.** Magnified view of the region surrounded by a red rectangle in **a**. on a y-scale expanded 22000-fold. The signal is due to the characteristic interferometric modulation induced by the molecular lines.

An interferogram of acetylene in the region of emission of the oscillators (182-202 THz) is shown in Fig. 2 over the entire range of optical delays of $1/f_{rep}$=10 ns. The interferogram results from 29000 averages over a total measurement time of 290 s. We can choose to average the individual interferograms in the time domain or the spectra in the frequency domain, the first approach has the advantage of a significantly reduced data file size with easier storage and shorter computation time. When we continuously average the interferograms in the time domain, we do not perform phase correction or any other operations on the raw data other than summing them. We then compute the complex spectrum by Fourier transformation and derive the phase and amplitude (Fig. 3a) of the resulting spectrum. For any averaging times, the signal-to-noise ratio in such time-domain-averaged spectrum is systematically 6% smaller than when we average the individual amplitude spectra.

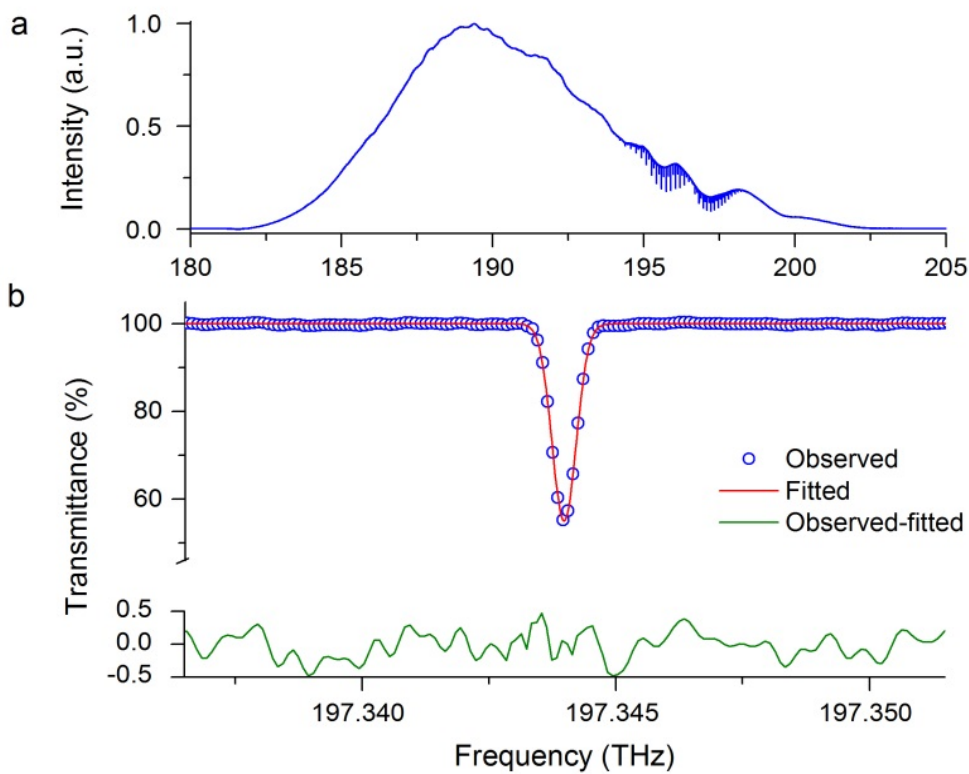

**Figure 3**. Spectrum in the region of the $v_1+v_3$ combination band of $^{12}C_2H_2$ with a resolution of 100 MHz corresponding exactly to the comb line spacing. **a**. The entire spectrum spans 20 THz and it is measured within 290 s. **b**. Magnified representation of a) showing the $R(11)$ line of the $v_1+v_3$ band of $^{12}C_2H_2$. The transmittance y-scale only goes down to 55%. A Doppler profile (red line) fits the experimental spectrum (blue dots) data is fitted with a Gaussian profile. The standard deviation of the residuals (green) between the experimental data and the fit is 0.18%.

A spectrum displaying resolved comb lines resulting from a measurement time of 209 s is shown in Fig. 4. Seven hundred interferograms, of 0.298 s duration each, have been averaged. In the spectrum, more than $2 \times 10^5$ individual comb lines are resolved across 20 THz. The observed full-width at half maximum of the comb lines is set by the measurement time of an individual interferogram to 4.0 Hz in the radio-frequency domain. Each comb line (Fig.4c) appears as a cardinal sine, the expected instrumental line shape in a non-apodized spectrum. In our previous report [3], it was washed out by residual phase fluctuations.

For a measurement time of 290 s (Fig. 3a), the signal-to-noise ratio in the spectrum culminates at 3180 around 189.5 THz and the average signal-to-noise ratio across the entire span of 20 THz is 1432. The resulting figure of merit, calculated for the average signal-to-noise ratio, is therefore $1.7 \times 10^7 \, \text{Hz}^{1/2}$. A limitation to the sensitivity is the need to restrict the power falling onto the detectors to avoid artifacts induced by the nonlinearities. Therefore, our value for the figure of merit is slightly higher, but of the same order of magnitude, than that reported in [4].

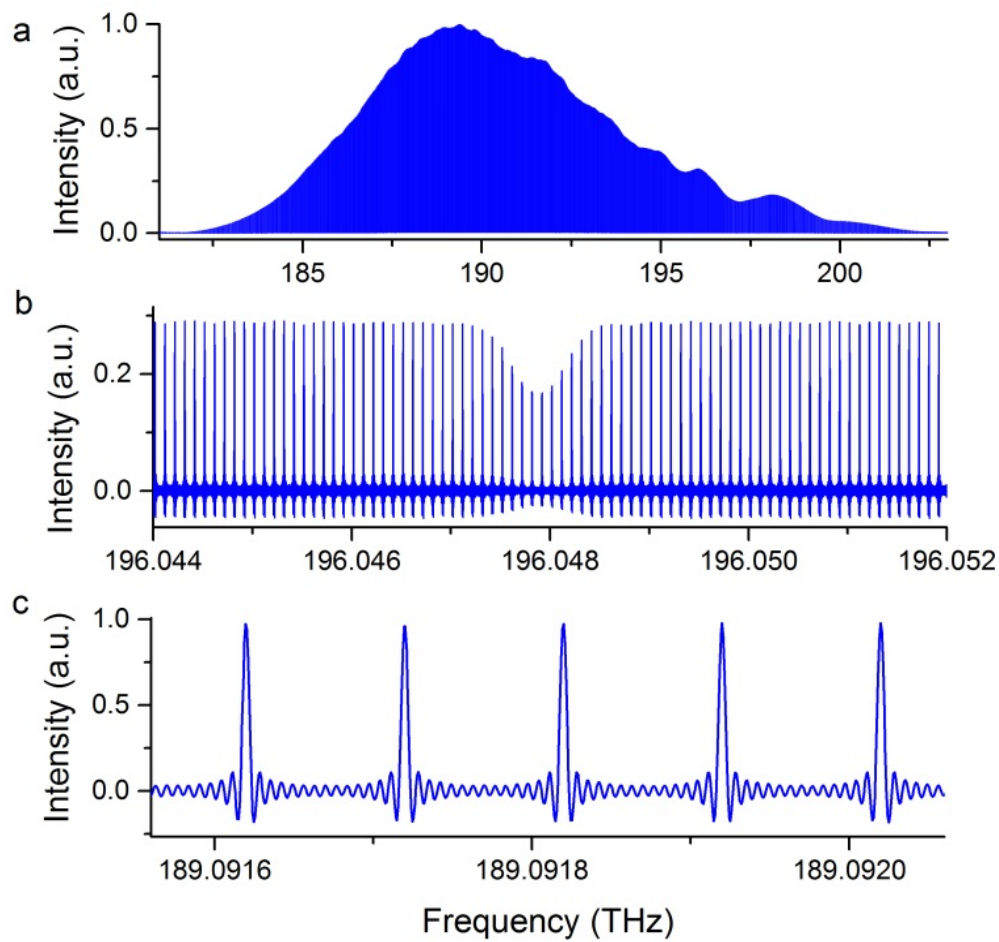

**Figure 4. Spectrum around 190 THz with resolved comb lines, measured within 209 s a**. Entire spectrum across the entire domain of emission of the laser oscillators: more than 200000 individual comb lines spanning 20 THz are resolved. **b**. Magnified representation of a. showing the $P(7)$ Doppler-broadened line of the $v_1+v_3$ band of $^{12}C_2H_2$ sampled by the comb lines of 100 MHz line spacing **c**. Magnified representation of a showing five individual comb lines. A cardinal sine instrumental line shape convolves the unapodized lines.

Figure 5 displays the evolution of the average signal-to-noise ratio with time (or number of averages). It increases with the square root of the measurement time, showing that the mutual coherence between the two combs is preserved throughout. The same behavior is observed around 180 THz and around 230 THz. The maximum averaging time that we report, 290 s, is a technical limitation due to our digitizer. No saturation in the trend of the increasing signal-to-noise ratio is observed though. This suggests that, with a dedicated data acquisition system, coherence over longer averaging times will be achieved.

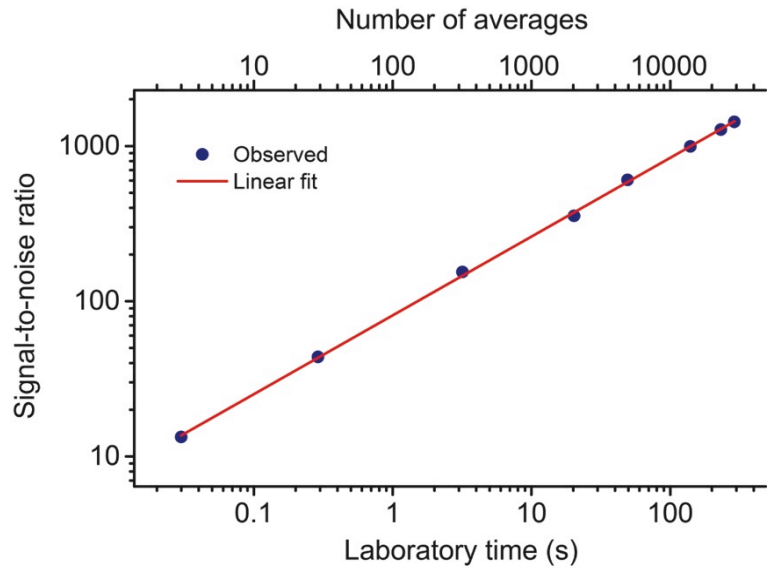

**Figure 5. Evolution of the signal-to-noise ratio in the spectra with the measurement time.** The interferograms are averaged in the time domain and Fourier transformed with a resolution of 100 MHz. The average signal-to-noise ratio in the spectra is plotted for different times of averaging. A linear fit with a slope of 0.507(4) * indicates that the signal-to-noise is proportional to the square root of the measurement time, which is expected for coherent averaging. The trend has not reached its limit at the maximum measurement time of 290 s.
* The number in parentheses is the standard deviation in units of the last quoted digit.

Good signal-to-noise ratios are achieved all across the region spectrally broadened by nonlinear fibers. As an example, a spectrum of methane in the region of the $2\nu_3$ band around 180 THz is shown (Fig.6a) at 100-MHz resolution. The filtered spectral span is 175-184 THz and the recording time is 14.46 s. The signal-to-noise ratio is at best 770 around 183.2 THz and the average signal-to-noise ratio is 465, which leads to a figure of merit of $1.1 \times 10^7$ $Hz^{1/2}$.

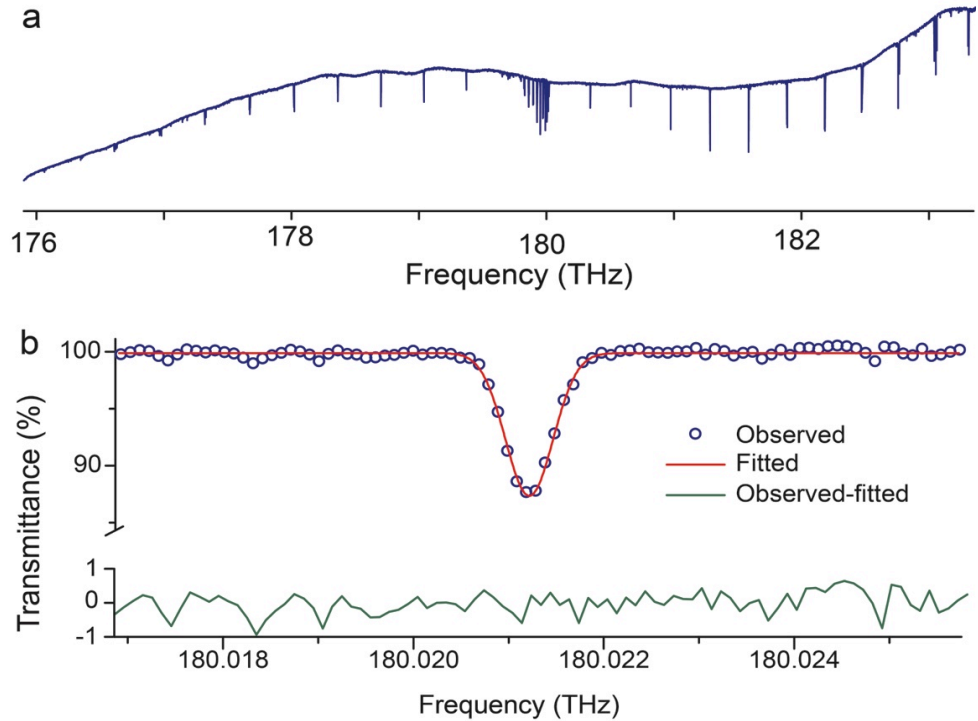

**Figure 6. Experimental dual-comb spectrum at a resolution of 100 MHz in the 180 THz region. a.** The spectrum of the $2\nu_3$ band of $^{12}CH_4$ spans 9 THz and it is measured within 14.46 s. **b**. Portion of the spectrum shown in a, magnifying the $Q(1)$ line with about 13% of absorption at line center. The experimental profile is satisfactorily fitted by a Doppler lineshape. The standard deviation of the residuals „observed – fitted" is 0.3%, at the noise level.

We determine the line positions in our self-calibrated spectra by fitting Doppler profiles of fixed width to the experimental transitions. Because of the long averaging time, of the narrow width of the optical comb lines and of the high mutual coherence of the dual-comb system, the instrumental line shape, that convolves the profiles, can be neglected. The residuals of the fit do not show any systematic signatures, as exemplified in Fig. 3b with the $R(11)$ line of the $\nu_1+\nu_3$ band of $^{12}C_2H_2$ and in Fig. 6b with the $Q(1)F_2$ line of the $2\nu_3$ band of $^{12}CH_4$. For the $\nu_1+\nu_3$ band of $^{12}C_2H_2$, we compare the positions of eleven of our experimental lines, for which the self-induced pressure shift has been measured in [24], to accurate sub-Doppler saturated absorption measurements [25,26]. The mean value of the discrepancies between our measurements and those of ref. [25] and ref. [26], respectively, is 120 kHz and 11 kHz, respectively, with a standard deviation of 350 kHz and 425 kHz, respectively. Our accuracy is primarily determined by the statistical uncertainty. Additional details may be found in Table 1.

With our demonstration of a dual-comb interferometer with long mutual-coherence times, new opportunities for broadband metrology are opened up. Moreover, our technique is expected to significantly improve the performance of dual-comb interferometers in spectral regions like the mid-infrared domain, where narrow-linewidth continuous-wave lasers are challenging to develop or where fast actuators for the two degrees of freedom of the combs are not straightforwardly available.

| Assignment | Frequency (MHz) | | | | | | |
|---|---|---|---|---|---|---|---|
| | This work | Pressure shift at 106.7 Pa [24] | This work extrapolated to zero pressure | [25] | [26] | This work-[25] | This work-[26] |
| P(14) | 195500510.5 (12) | -0.22(2) | 195500510.7(12) | 195500510.62(15) | 195500510.7477(101) | 0.1 | 0.0 |
| P(13) | 195580978.7 (7) | -0.22(2) | 195580978.9(7) | 195580979.28(13) | 195580979.3711(102) | -0.4 | -0.5 |
| P(10) | 195817848.5 (11) | -0.23(2) | 195817848.7(11) | 195817848.23(15) | 195817848.3823(105) | 0.5 | 0.3 |
| P(6) | 196123038.4 (10) | -0.22(4) | 196123038.6(10) | 196123038.51(15) | 196123038.5214(103) | 0.1 | 0.1 |
| P(5) | 196197428.3 (10) | -0.23(2) | 196197428.5(10) | 196197428.20(13) | 196197428.3457(102) | 0.3 | 0.2 |
| P(4) | 196271052.8 (9) | -0.23(2) | 196271053.0(9) | 196271052.46(14) | 196271052.5819(103) | 0.5 | 0.4 |
| P(3) | 196343910.4 (8) | -0.22(2) | 196343910.6(8) | 196343910.06(14) | 196343910.0006(127) | 0.5 | 0.6 |
| R(1) | 196696652.2 (10) | -0.11(2) | 196696652.3(10) | - | 196696652.9203(100) | - | -0.6 |
| R(7) | 197094394.4 (10) | -0.23(2) | 197094394.6(10) | 197094394.83(15) | 197094395.0333(104) | -0.2 | -0.4 |
| R(11) | 197343961.7 (7) | -0.31(2) | 197343962.0(7) | 197343962.37(15) | - | -0.4 | - |
| R(17) | 197694759.9 (12) | -0.44(2) | 197694760.3(12) | 197694760.15(15) | - | 0.2 | - |

**Table 1.** Center frequencies of the Doppler-broadened lines in the $v_1+v_3$ band of $^{12}C_2H_2$ measured in this work (retrieved from the spectrum shown in Figure 3 measured within 290s) along with a comparison with the absolute frequency measurements reported in [25] and [26] by Doppler-free saturated absorption spectroscopy. Our experimental line positions at zero pressure are corrected (fourth column) for the pressure shift using the values measured by [24]. We only provide frequencies for the lines for which the pressure shift had been determined and accurate measurements for comparisons are available. The number within parentheses is the uncertainty, including statistical and systematic effects, in units of the last digit.

**Acknowledgments.** We warmly thank Dr. Gwénaëlle Mélen for developing the program to compute the phase and amplitude of the spectra with resolved comb lines. We gratefully acknowledge funding from the European Research Council (Advanced Investigator Grant 267854), the Munich Center for Advanced Photonics and the Max-Planck Foundation.